\documentclass[ twocolumn]{IEEEtran}
\usepackage{times}
\usepackage{amsmath,epsfig}
\usepackage{amssymb}
\usepackage{amsfonts}
\usepackage{array,dcolumn}
\usepackage[ruled,vlined,linesnumbered]{algorithm2e}
\usepackage{subcaption}
\usepackage{psfrag}
\usepackage{amsmath}
\usepackage{amsfonts}
\usepackage{amssymb}
\usepackage{balance}
\usepackage{accents}
\usepackage{booktabs, adjustbox}
\usepackage{cite}
\usepackage[all]{xy}
\usepackage{dsfont}
\usepackage{hyperref}
\usepackage{url}
\usepackage{xcolor}
\usepackage[nolist]{acronym}
\usepackage{graphicx}
\usepackage{threeparttable}
\usepackage{cancel}

\hypersetup{
     colorlinks = true,
     linkcolor = black,
     anchorcolor = black,
     citecolor = black,
     filecolor = blue,
     urlcolor = blue
     }

\newcommand{\LN}[1]{\textcolor{black}{#1}}
\newcommand{\LD}[1]{\textcolor{black}{#1}}
\newcommand{\PP}[1]{\textcolor{black}{#1}}
\newcommand{\IL}{\textcolor{black}}
\newcommand{\rev}{\textcolor{black}}
\newcommand{\IM}{\textcolor{black}}

\title{Trusted Wireless Monitoring based on Distributed Ledgers over NB-IoT Connectivity}

\author{\IEEEauthorblockN{Lam D. Nguyen, \textit{Student Member, IEEE}, Anders E. Kal{\o}r, \textit{Student Member, IEEE}, \\ Israel Leyva-Mayorga, \textit{Member, IEEE}, and Petar Popovski, \textit{Fellow, IEEE}}\\
\thanks{The authors are with Connectivity Section, Department of Electronic Systems, Aalborg University, Aalborg, Denmark.
E-mail: \{ndl, aek, ilm, petarp\}@es.aau.dk}
}
\begin{document}
\maketitle

\begin{abstract}
    \PP{The data collected from  Internet of Things (IoT) devices on various emissions or pollution, can have a significant economic value for the stakeholders. This makes it prone to abuse or tampering and brings forward the need to integrate IoT with  a Distributed Ledger Technology (DLT) to collect, store, and protect the IoT data.} 
    However, DLT brings an additional overhead to the frugal IoT connectivity and symmetrizes the IoT traffic, thus changing the usual assumption that IoT is uplink-oriented. We  have implemented a platform that 
    integrates DLTs with a monitoring system based on narrowband IoT (NB-IoT). We evaluate the performance and discuss the tradeoffs in two use cases: data authorization and real-time monitoring.
\end{abstract}

\begin{IEEEkeywords}
Distributed Ledger Technologies, Blockchain, Transparency, Narrowband Internet of Things, Environment Monitoring, Air Pollution.
\end{IEEEkeywords}

\section{Introduction}
\label{sec:introduction}

\PP{An important element in the process of combating climate change and protecting public health is the reliable and trustworthy measurement of various emissions and air pollutants. Prime examples include CO\textsubscript{2} and NO\textsubscript{x}, for which monitoring systems based on Internet of Things (IoT) technology have been reported in~\cite{abawajy2017federated}.
The emission information is critical and can have a significant economic value, such that the stakeholders have incentives to manipulate the data. The way this information from IoT-based monitoring systems is stored and collected raises concerns about data integrity, trust, security, transparency, and public availability.}
For instance, in IoT deployments, the measured data are either centralized or spread out across different heterogeneous parties. These data can be both public or private, which makes it difficult to validate their origin and consistency. Besides, querying and performing operations on the data becomes a challenge due to the incompatibility between different application programming interfaces (APIs). For instance, Non-Governmental Organizations (NGOs), Public and Private sectors, and industrial companies may use different data types and databases, which leads to difficulties when sharing the data. 

Data authorization represents another critical component in many monitoring applications, in which the validity of the received information is critical. To this end, IoT monitoring systems often rely on an intermediary entity to validate the device signatures, e.g., a certificate authority (CA) server, which suffers from the issue of a single point of failure. As a result, the data from authenticated devices are vulnerable to tampering using, for example, man-in-the-middle attacks against the CA server. 

\IL{Distributed ledger technologies (DLTs) are positioned as a key enabler for trusted and reliable distributed monitoring systems, since these support the immutable and transparent information sharing among involved untrusted parties~\cite{previous}. In DLTs, the authentication process relies on consensus among multiple nodes in the network. While the terms DLT and \emph{Blockchain} will be used interchangeably throughout this paper, Blockchains are a type of DLT, where chains of blocks are made up of digital pieces of information called transactions and every node maintains a copy of the ledger. Therefore, in a Blockchain-enabled IoT network, transactions contain, for example, environmental sensing data, or monitoring control messages, and these are recorded and synchronized in a distributed manner in all the participants of the system. These participants are called miners or peers and, in some specific DLTs, users are charged a transaction fee to perform (crypto) transactions.} In addition, DLTs allow the storage of all transaction into immutable records and every record distributed across many participants. Thus, security in DLTs comes from the distributed characteristic, but also the use of strong public-key cryptography and strong cryptographic hashes.

\begin{figure*}[t]
    \centering
    \includegraphics[width=0.85\linewidth]{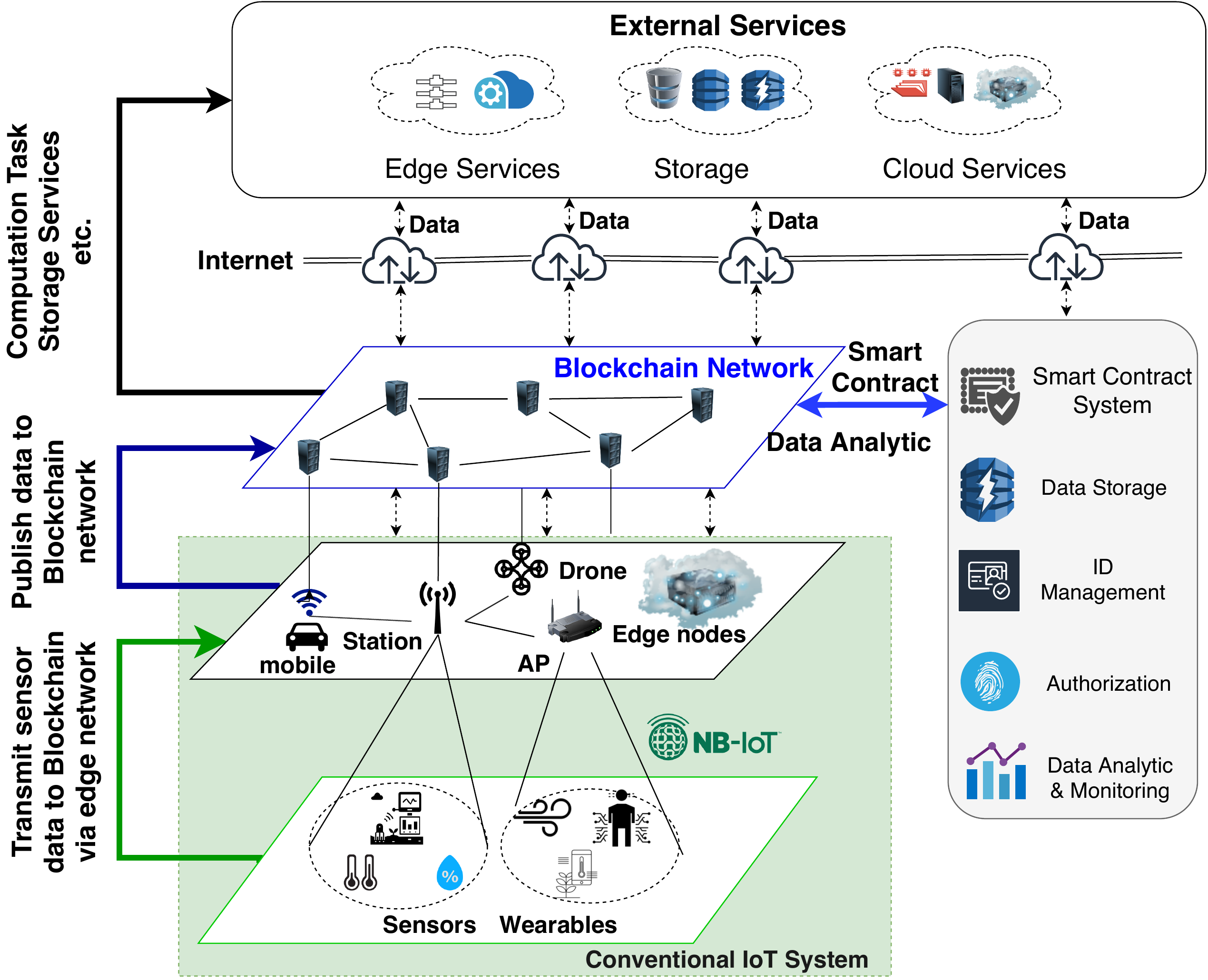}
    \caption{General DLT-enabled NB-IoT pollution monitoring architecture}
    \label{fig:architecture}
\end{figure*}


\PP{The benefits of the integration of DLTs into IoT monitoring systems include: i) guarantee of immutability and transparency for environmental sensing data; ii) removal of the need for third parties; iii) development of a transparent system for heterogeneous IoT monitoring networks to prevent tampering and injection of fake data from the stakeholders.}

In this article, we describe and analyze the tradeoffs of the integration of DLTs into the narrowband Internet of Things (NB-IoT), which currently is the leading cellular IoT technology~\cite{wang2017primer}. NB-IoT is one of the most efficient low-power wide-area network (LPWAN) technologies in terms of coverage, battery life-time, and support for massive machine-to-machine communications (i.e., scalability)~\cite{azari2019latency}. \rev{A feature that makes NB-IoT more suitable than other technologies to support DLT traffic is its high downlink and uplink capacity when compared to other LPWANs such as LoRaWAN or Sigfox ~\cite{mekki2019comparative}. For instance, the suitability of LoRaWAN to support DLT traffic is mainly limited by its modest data rates and its $1$ percent duty cycling (i.e., nodes must be idle $99$~percent of the time)~\cite{previous}. Conversely,} NB-IoT has been designed with adaptable data rates and high flexibility, bringing significant advantages to sensing and monitoring networks. For instance, NB-IoT can be configured to use a wide range of \IL{sub-carrier spacing} settings, which allows the protocol to be tailored for the specific deployment scenario and data rates can be increased 12 times by allocating multiple sub-carriers to the devices. NB-IoT provides extended coverage low-power devices with battery life-time up to 15 years\cite{wang2017primer}. Besides, NB-IoT is optimized for regular and small data transmissions, so it is well suited for monitoring devices acting as air quality, gas, and water meters~\cite{feltrin2019narrowband}. 

We aim for a full integration where the NB-IoT devices generate transactions and receive the corresponding confirmations, but do not act as Blockchain nodes. Such integration is analogous to the P2 protocol described by Danzi \emph{et. al}~\cite{Danzi2018}, which provides end-to-end (E2E) security and trust without increasing the storage and computation load of IoT devices. On the downside, such integration raises the following questions: i) how does Blockchain consensus and synchronization affect the NB-IoT connectivity in terms of uplink and downlink traffic and end-to-end (E2E) latency? and ii) which trade-offs arise from integrating DLTs into NB-IoT monitoring systems? For instance, the traffic patterns generated by DLTs are different to traditional IoT traffic, where the ratio of uplink (UL) to downlink (DL) data is oftentimes small. That is, most of the data is usually transmitted from the NB-IoT devices to the network to be stored and processed. Instead, the need to maintain a ledger for all the participants increases the amount of data transmitted from the base station (i.e., DL).

To answer these above questions, we first describe in detail the essential elements of an integrated Blockchain and NB-IoT system that addresses the problem of trust and privacy (Section \ref{sec:architecture}). Then, we discuss and analyze the relevant characteristics of popular DLTs platforms such as Bitcoin, Ethereum, Hyperledger Fabric, and IoTA, and select the most promising to integrate into IoT environment monitoring systems. Additionally, we provide an overview of the operation of NB-IoT.
Finally, we investigate the suitability of NB-IoT to connect the physical monitoring system with the Blockchain in two specific use cases, namely data authorization and real-time \rev{(i.e., timely)} monitoring of gas emissions (Section \ref{sec:cases}). In particular, we analyze and evaluate the effect of Blockchain in NB-IoT monitoring systems in terms of traffic balance (DL to UL), communication overhead, and E2E latency, measured as the transaction confirmation time.

Our results, obtained from extensive experiments, show that the mining and consensus mechanisms allow Blockchain nodes to reach a secure and tamper-resistant consensus in collected sensing data. On the downside, we observed an increase in the amount of DL data and E2E latency, in the order of a few seconds, when compared to traditional NB-IoT packet transmissions.
Despite these minor drawbacks, we consider that Blockchain and NB-IoT can have a symbiotic relationship to provide data integrity, trust, security, transparency, and public availability for a wide range of monitoring systems with a minimal impact on energy-efficiency.

\rev{In particular, the contributions of this work are threefold. First, we present a Blockchain-powered IoT framework for environmental monitoring systems that addresses the problem of trust and privacy. Second, we evaluate the proposed framework via extensive experiments, in which the NB-IoT monitoring system and a suitable DLT platform are integrated. Third, realizing the lack of studies on communication aspects of current Blockchain-enabled IoT systems, we analyze and evaluate the interaction between Blockchain and the NB-IoT monitoring systems in terms of overall throughput, E2E latency, and communication overhead via two case studies. Regarding previous studies \cite{dai2019blockchain}, to our best knowledge, these studies mainly focus on specific-applications of Blockchain-enabled IoT, and how to integrate Blockchain with IoT. In our studies, communication aspects between Blockchain nodes and IoT devices are investigated.}


\section{Blockchain-powered IoT monitoring systems}
In this section, we describe the essential architectural elements for \IL{integration of Blockchain into NB-IoT monitoring systems}, evaluate the numerous DLT alternatives, and give a brief overview of the operation of NB-IoT.

\subsection{\rev{Essential architectural elements}} 
\label{sec:architecture}
The overall integrated system consists of 4 key components\IL{: DLT network, physical sensors, edge network, and external resources, as illustrated in Fig.~\ref{fig:architecture}. These components are described in following.}

\textbf{DLT Network:} This component includes all modules to build various features of Blockchain technologies such as consensus, smart contract, data authorization, identity management, and peer-to-peer (P2P) communication. These components must ensure that every change to the ledger is reflected in all copies in seconds or minutes and provide mechanisms for the secure storage of the data generated by IoT devices and parameter configurations. There are numerous DLTs with different characteristics that may be beneficial for different target applications. \rev{The  DLT nodes can be located everywhere and connected with NB-IoT base stations via the Internet.}

\textbf{Physical Sensors:} The set of resource-limited devices which have the responsibility of collecting environmental data such as temperature, humidity, gas emissions and air quality levels. The collected data are transmitted to edge nodes or base stations, which can be static, such as access points, gateways, or mobile terminals, such as drones and mobile devices.

\textbf{Edge Network:} Even though DLT-based solutions offer significant countermeasures to secure data from tampering and support the distributed nature of the IoT, the massive amount of generated data from sensors and the high energy consumption required to verify transactions make these procedures unsuitable to execute directly on resource-limited IoT devices. Instead, edge servers with high computation resources can be used to handle real-time applications and to further increase the degree of privacy (e.g., through cloud computing)~\cite{xiong2018mobile}. The edge network is a potential entity to cooperate with the Blockchain network in computationally heavy tasks and return the estimation results \LN{(e.g., from solving proof-of-work (PoW) puzzles, hashing or algorithm encryption)} to the Blockchain network for verification. 

\textbf{External Services:}  IoT physical devices are resource-constrained with limited storage space and low computation capacity. Hence, external infrastructure may be incorporated to provide external services such as storage and computing. For example, the Interplanetary File System (IPFS) is a distributed file storage system that can store data generated from IoT networks and return a hash to the ledger based on the content of the data. Since the ledger cannot handle and store the massive amount of environmental data collected by the sensors, the services provided by the IPFS are a vital component.

\subsection{ \rev{Suitability of different DLTs for IoT monitoring}} 
\label{sec:dlt}
\begin{figure}[t]
    \centering
    \includegraphics[width=0.95\linewidth]{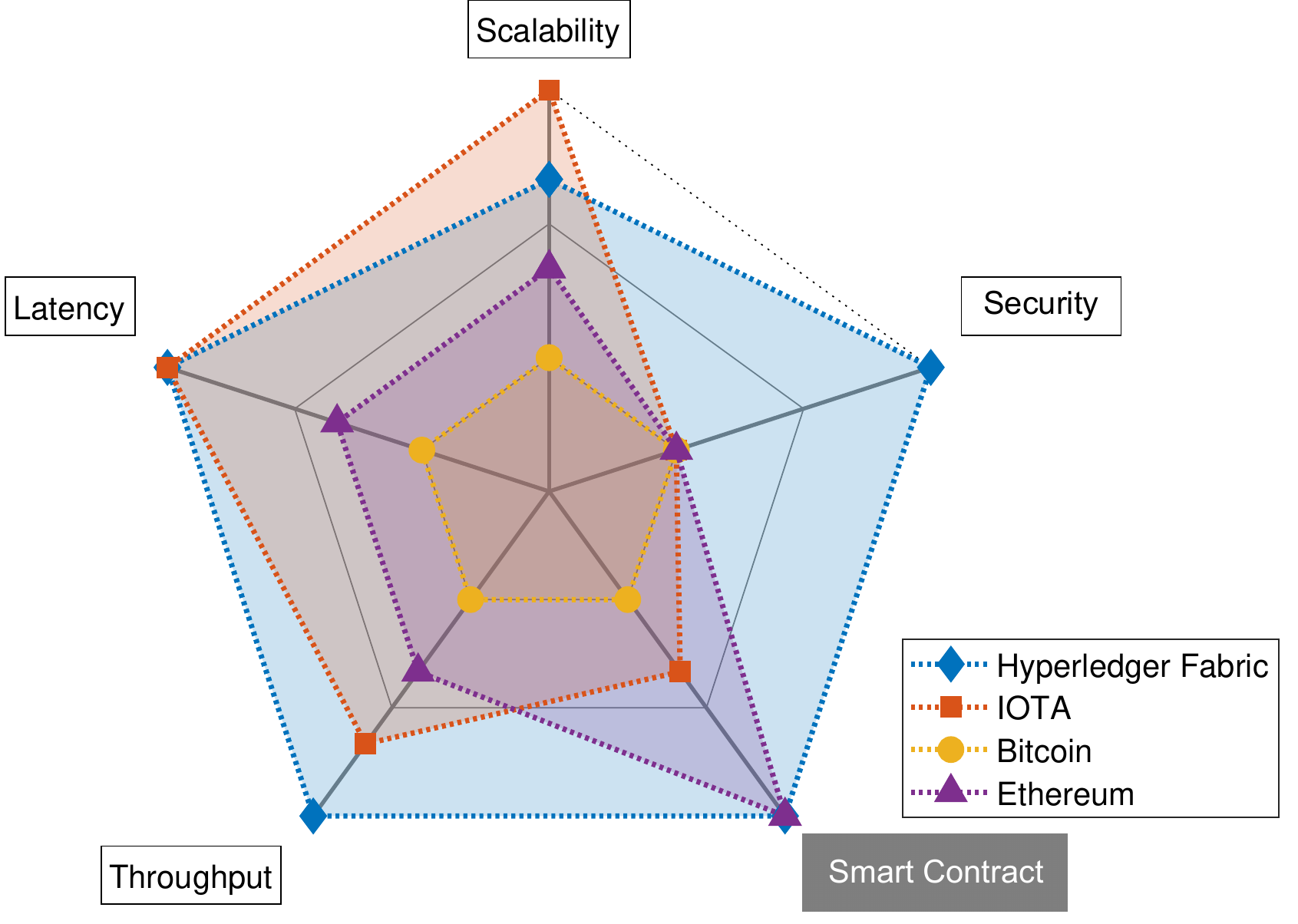}
    \caption{
    Performance of four different DLTs in five essential aspects for IoT monitoring systems dealing with sensitive information. }
    \label{fig:comparison}
\end{figure}%

Although a large number of Blockchain DLTs are available, the most prominent platforms include Bitcoin, Ethereum, IOTA, and Hyperledger Fabric. In the following, we compare these DLTs in five different aspects: scalability, latency, throughput, security, and the level of smart contract functionalities.

Scalability, latency, and throughput are deeply related and of vital importance for  IoT applications. For instance, the large number of sensors in smart cities may generate millions of transactions per day.
\PP{This requires high efficiency of the consensus mechanism, including the way in which transactions are processed by the peers, known as \emph{endorsing peers} in Hyperledger Fabric and full nodes (peers) in Bitcoin and Ethereum.} Regarding latency, the transaction confirmation time must be sufficiently short to avoid queueing in the Blockchain and to ensure consistency in the ledgers. Bitcoin and Ethereum confirmation times per transaction are around $10$ minutes and $25$ seconds, respectively. These latencies might not be suitable for real-time IoT monitoring, while the confirmation time of Fabric and IOTA is much lower \cite{xiao2020survey}. \LD{Note that the transaction confirmation time is only part of end-to-end latency, as it does not account for the communication latency at the radio access network.} 

The charge of fees to process the transactions, commonly known as gas is yet another factor to take into account to select the appropriate DLT. These may greatly increase the operational costs of the network, which negatively impacts the throughput of the DLT. On the one hand, transaction fees pose a problem in massive IoT scenarios if the generation of a large number of transactions is essential. On the other hand, these fees may contribute to minimize the amount of redundant transactions generated by the sensors, which in turn offloads the Blockchain. Among the considered DLTs, Ethereum requires fee and gas for each transaction whereas Hyperledger Fabric and IOTA provide free solutions to exchange transactions.

It is clear that IoT applications will involve many stakeholders with different roles, functionalities, and information with access rules, identities and security factors. An important factor to provide security is the support for permissioned and permissionless (i.e., hybrid) solutions to validate participating nodes. Both Ethereum and Hyperledger Fabric support public and private solutions, while Bitcoin and IOTA only provide public ones. Although IoT networks, such as smart cities, may have a large number of stakeholders willing to contribute to the security of a permissionless Blockchain network, permissioned networks could also be beneficial. For example, in smart homes where the homeowner wants to validate the transactions via home miners or validators\cite{lin2016iot}. \LD{Regarding security, public networks may be more secure than private ones if these are able to provide transparency and distributed storage. For instance, in a permissionless Blockchain, the data is encrypted and stored in all the devices, which makes it definitely transparent. Besides, the more users a permisionless Blockchain has, the more secure it is. However, permissionless Blockchains are not ideal for enterprise use, where companies deal with highly sensitive data and cannot allow anyone join their network. A permissioned Blockchain can be altered by its owners, making it more vulnerable to hacking\cite{compare}. In addition, permissioned Blockchains provide very low or no fee for validation and a faster consensus process.}

Finally, smart contracts act as autonomous entities on the ledger that deterministically execute logic expressed as functions of the data that are written on the ledger. Therefore, smart contracts can be established to have automatic reactions from the DLT network to specific events. \PP{For example, in case of carbon emissions, smart contracts can be used for real-time policy enforcement upon changes in the emission patterns.} The smart contract feature currently is supported by Ethereum and Hyperledger Fabric (Chaincode). An IOTA smart contract type called Quobic is still in progress. Besides, only Hyperledger Fabric supports data confidentially via in-band encryption and guarantees the privacy of data by creating private channels. Hyperledger Fabric provides a solution with various features such as identity management, transaction integrity and authorization with a trusted CA. These features are vital in a trusted IoT system. The comparison of the DLTs mentioned above in these areas illustrated in Fig.~\ref{fig:comparison}, where each aspect has been given an abstract score based on the previous discussion. \rev{Note that the smart contract aspect is a functionality rather than a strict performance indicator and can only be scored qualitatively. This makes it different to the rest of the aspects reflected in Fig.~\ref{fig:comparison}, hence, it is shown in a gray background.}

Based on these scores, we decided to implement Hyperledger Fabric as DLT platform for our experiments on IoT monitoring.

\subsection{DLT traffic over NB-IoT}
\label{sec:bcnbiot}
\begin{figure}[t]
    \centering
    \includegraphics[width=0.9\linewidth]{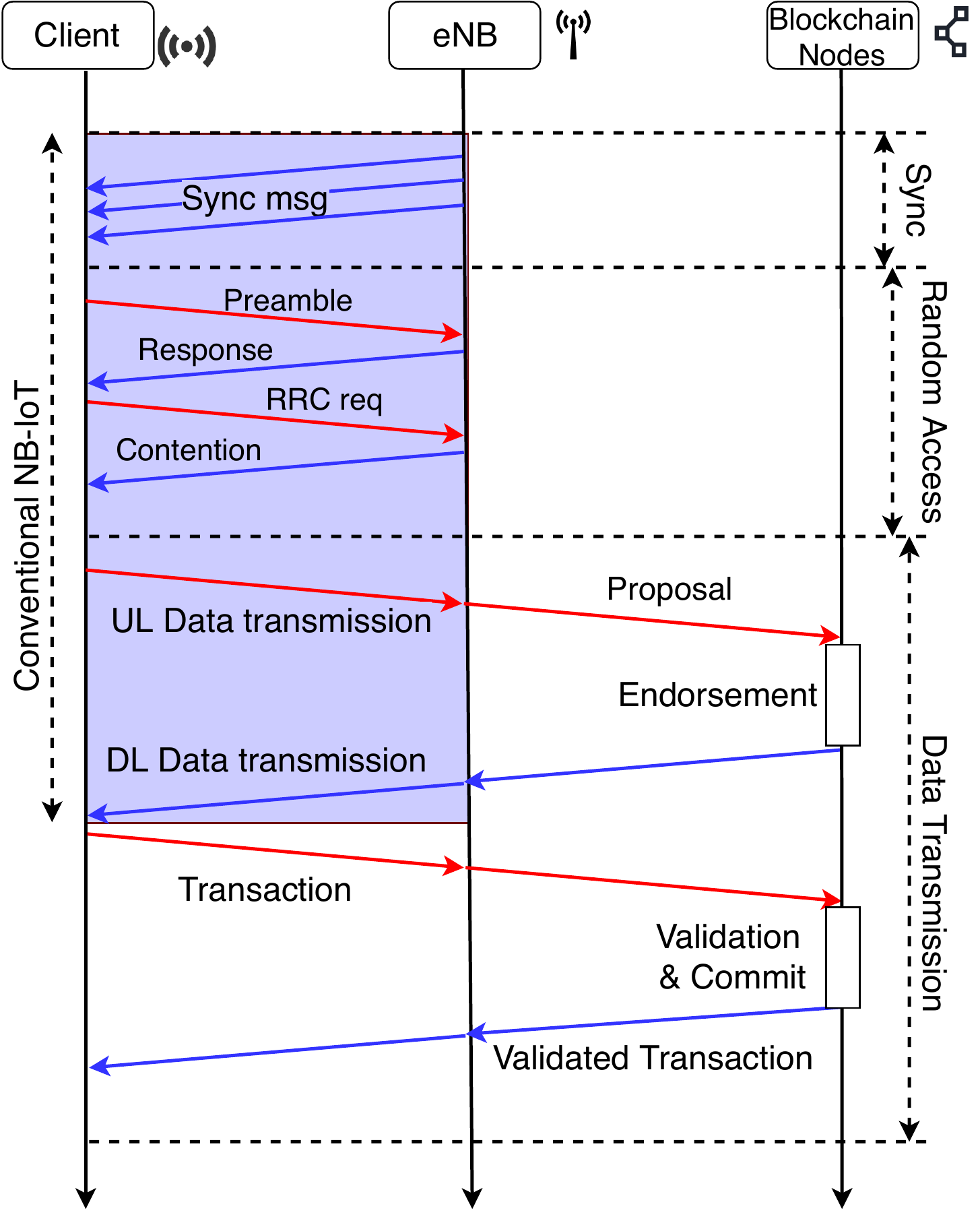}
    \caption{A sequence of message exchanges between DLT with UEs and eNB.} 
    \label{fig:workflow}
\end{figure}%

In the following, we provide a brief description on the operation of NB-IoT devices, hereafter referred to as user equipments (UEs), in monitoring applications. \IL{NB-IoT UEs have only two modes of operation, namely radio-resource control (RRC) idle and RRC connected. In the former, the UEs can only receive the system information from the BS and, only in the latter, data can be transmitted. UEs are in idle mode before initial access to the network, but may also enter this mode during power saving or after an explicit disconnection request. To transition from idle to connected mode, the UEs (clients) must first acquire the basic system information and synchronization as illustrated in the upper part of Fig.~\ref{fig:workflow}. For this, the UE receives the master information block (MIB-NB) and the system information blocks 1 (SIB1-NB) and 2 (SIB2-NB). These are transmitted periodically through the downlink shared channel (DL-SCH) and carry the basic cell configuration, timing, and access parameters~\cite{3GPPTS36331}. In addition, SIB1-NB carries the scheduling information for the rest of the SIBs.} 

\IL{After the system information has been acquired, the UEs must perform the RA procedure to transition to RRC connected mode~\cite{3GPPTS36321}. The RA procedure is a four-message handshake, initiated by the UEs by transmitting a single-tone frequency-hopping pattern, called preamble, through the NB Physical Random Access Channel (NPRACH). In most cases, the RA procedure is contention-based, hence, the preamble is chosen randomly from a predefined pool of up to $48$ orthogonal sub-carrier frequencies. Consequently, the main reasons for an access failure are the lack of power in the transmission and simultaneous transmissions of the same preamble, which lead to collisions.}

After completing the RA procedure, and if the control-plane (CP) cellular IoT (CIoT) is used, UEs may piggyback short UL data packets along with the RRC Connection Setup Complete message. Otherwise, the non-access stratum (NAS) setup must be completed before eNB allocates resources for uplink transmission through the NB Physical UL Shared Channel (NPUSCH) and data can be transmitted. \IL{The resource unit (RU) is the basic unit for resource allocation in the NPUSCH and comprises a set of sub-frames in the time domain and sub-carriers in the frequency domain. The downlink (DL) data is transmitted through the NB physical DL shared channel (PDSCH).}

\IL{In a traditional NB-IoT monitoring system, the UL data generated by the UEs is transmitted though the NPUSCH and routed towards a data center or cloud server to be stored and processed. At this point, the monitoring system has no control on the collected data, so modification, corruption, and losses may occur.} \IL{Conversely, in our Blockchain-enabled NB-IoT setup,} the uplink data generated by the UEs is transmitted to a randomly chosen group of endorsing peers of Hyperledger Fabric as transaction proposals. Then, each of the peers signs the transaction using Elliptic Curve Digital Signature Algorithm (ECDSA) and adds the signature before returning the signed message back to the UEs.    
The peers that provide an endorsement of the proposed ledger send an update to the application, but do not immediately apply the proposed update to their copy of the ledger. Instead, a response is sent back to the UEs to confirm that the transaction proposal is correct, has not been previously submitted to ledger, and has a valid signature. Therefore, the security increases with the number of endorsing peers.
In addition, smart contracts can be executed to update or query the ledger. \IM{A simple example of a smart contract in air pollution monitoring systems would be to set the system to calculate average values of the collected data and to generate an alarm message whenever these exceed a predefined threshold.}

Then, the UEs broadcast the confirmed transaction proposals along with the response (confirmation) to the ordering service. \IL{The received transactions are ordered chronologically to create blocks}. These blocks of transactions are delivered to all the peers for validation. The peers append the block to the ledger, and the valid transactions are committed to the current state database. Finally, a confirmation message is emitted and transmitted back to the UEs to notify that the submitted transaction has been immutably published to the ledger. \rev{ In our previous work\cite{previous}, we have shown that two-way wireless communication is required to enable high decentralization. In communication aspects, we studied and analyzed DLT traffic in both uplink and downlink over IoT networks. The confirmation serves as proof that a transaction executed and recorded in distributed ledger.  This confirmation helps to check for errors or strange occurrences, for example, in retail industry, the confirmation support to detect the number of items re-bought, etc. Then, it is fed into the overall retail system, inventory system, and more are updated. Furthermore, if the sensors do not receive confirmation from the distributed ledger, the next action depends on specific applications and configuration. We can configure the NB-IoT sensor just to collect sensing information and publish to the distributed ledger without waiting for receiving any feedback. However, it is a trade-off between overheads and security. In reality, the confirmation message might include not only the confirmation from the distributed ledger to sensor but also execution commands generated autonomously defined by smart contracts.}

\begin{figure}[t]
    \centering
    \includegraphics[width=1.0\linewidth]{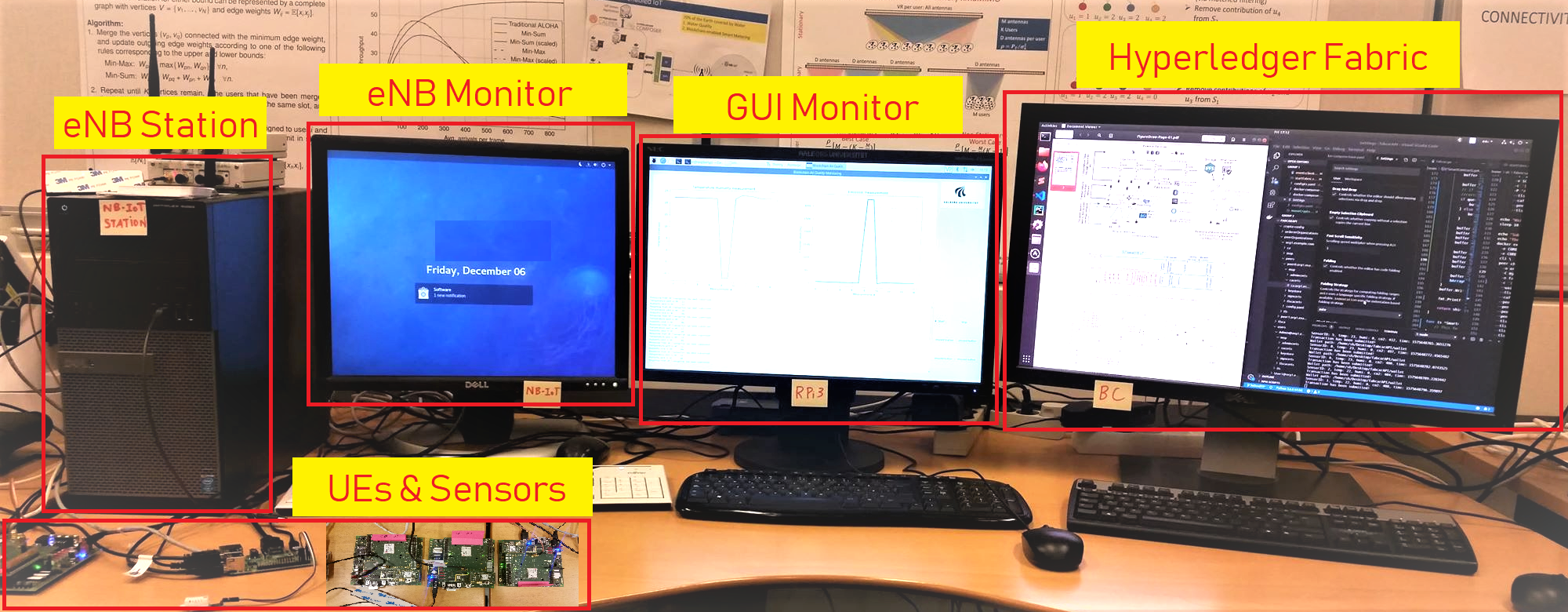}
    \caption{Blockchain-enabled NB-IoT implementation and setup.}
    \label{fig:setup}
\end{figure}

\section{Case Studies}
\label{sec:cases}
In this section, we evaluate the performance of an integrated Blockchain and IoT monitoring system with Hyperledger Fabric and NB-IoT under two use cases. The first one focuses on the data authentication aspect provided by Blockchain. We then extend the use case to include smart contracts to processes the sensor measurements. Our experimental setup is based on Hyperledger Fabric v1.4, NB-IoT development kits Sara EVK N211, and one NB-IoT Amarisoft eNB station, and is illustrated in Fig.~\ref{fig:setup}.

\subsection{Use case 1: Data Authorization}
\label{sec:cs1}

The focus of this use case is to evaluate the communication overhead of data authentication in our setup. Our setup includes a single UE with a single sensor that follows the procedure described in Section~\ref{sec:bcnbiot}; illustrated in Fig.~\ref{fig:workflow}. The metric we use to evaluate the communication efficiency is the average UL to DL data traffic ratio, where only data packets are considered. 

The size of the payload in the transmitted packets plays a vital role in the performance of DLT-based NB-IoT systems. Therefore, we varied the UL payload size from $50$~B to $200$~B and set the UE to generate a total of $1000$ transactions (i.e., UL data packets). The DL payload size is set to $31$~B, so the UL payload size is at least $1.61$ times the DL payload size, and the block size is configured to $30$ transactions per block. We ran our experiments with different number  of endorsing peers in the DLT network $E$ to observe the communication overhead of an increase in security, which, naturally, increases with $E$. 

\begin{figure}[t]
    \centering
    \includegraphics[width=0.9\linewidth]{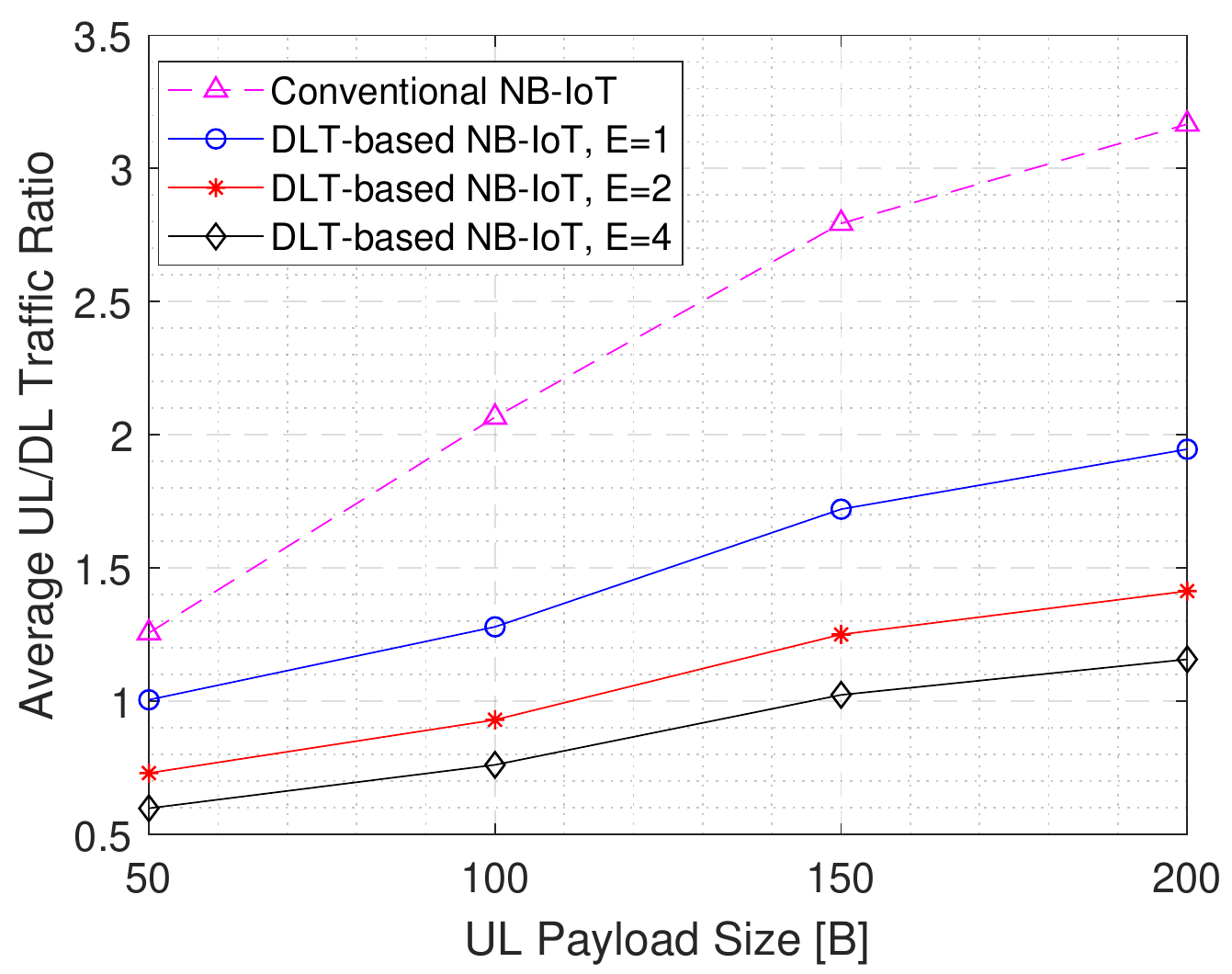}
    \caption{Average ratio of UL and DL traffic per transaction with different payload sizes and numbers of endorsing peers $E$. }
    \label{fig:throughput}
\end{figure}

Our results are presented in Fig.~\ref{fig:throughput}, where it can be seen that, naturally, the average UL to DL traffic ratio increases with the UL payload size. However, this increase is more rapid with traditional NB-IoT than with Blockchain and the average UL to DL traffic ratio decreases as $E$ increases. For instance, with an UL payload size of $50$~B, and $E=2$, the DL traffic is almost twice as high as the UL traffic. The reason for this is that, certainly, the average traffic increases with the number of endorsing peers \IL{$E$}, but the increase in DL traffic is much greater than the increase in UL traffic. Hence, these results highlight the fact that the use of DLTs heavily increases the traffic load in the DL channels of IoT networks, namely, in the PDSCH of NB-IoT.

\subsection{Use case 2: Real-time Monitoring of Air Pollution}
\label{sec:cs2}

We now study the specific use case of a real-time $\text{CO}_2$ emission monitoring system that includes smart contracts in the DLT. The environment data is collected by $S8$ Miniature $10000$~ppm $\text{CO}_2$~sensors in the UEs; a simple smart contract is defined to compute the average $\text{CO}_2$ level and trigger updates to the ledgers when these levels are abnormally high.

\rev{The environment data is collected once every 10 seconds by S8 Miniature 10000 ppm CO 2 sensors in the UEs; a simple smart contract is defined to compute the average indoor levels of CO2 and to trigger updates to the ledgers when these levels are abnormally high. Note that high indoor CO2 levels are greatly correlated to human metabolic activity and can cause headaches or make the population to function at lower activity levels. Nevertheless, the CO2 emissions generated by working equipment (e.g., computers, machines, etc.), also have an impact on total amount of CO2 emissions. Our experiments on CO2 and NOx data were conducted using the same methods to collect and process the data. Hence, the type of sensor does not affect the generated traffic and system process.}

The focus in this use case is to evaluate the E2E latency, defined as the time elapsed from the generation of a transaction at the IoT device until its verification. This includes the latency at the NB-IoT radio link and at the DLT, which comprises the execution time of the smart contract and transaction verification. Therefore, the E2E latency of smart contract execution depends on the numerous parameters such as block size and transaction generation rate. Among these, we evaluate the impact of the block size on E2E latency. \rev{Our results indicate that integrating DLTs into NB-IoT monitoring applications symmetrizes the data traffic by slightly increasing the amount of data transmitted in the downlink. However, by adequately choosing the DLT and its parameters, the impact of DLT traffic on the battery lifetime of NB-IoT nodes may be relatively low when compared to that of the traffic pattern of the monitoring application and of the implemented power saving techniques. Hence, our results can be combined with detailed energy consumption models that include the different possible states and power saving techniques of NB-IoT nodes (e.g., \cite{lauridsen2018empirical}) to estimate their battery lifetime.
}

\begin{figure}[t]
    \centering
    \includegraphics[width=0.9 \linewidth]{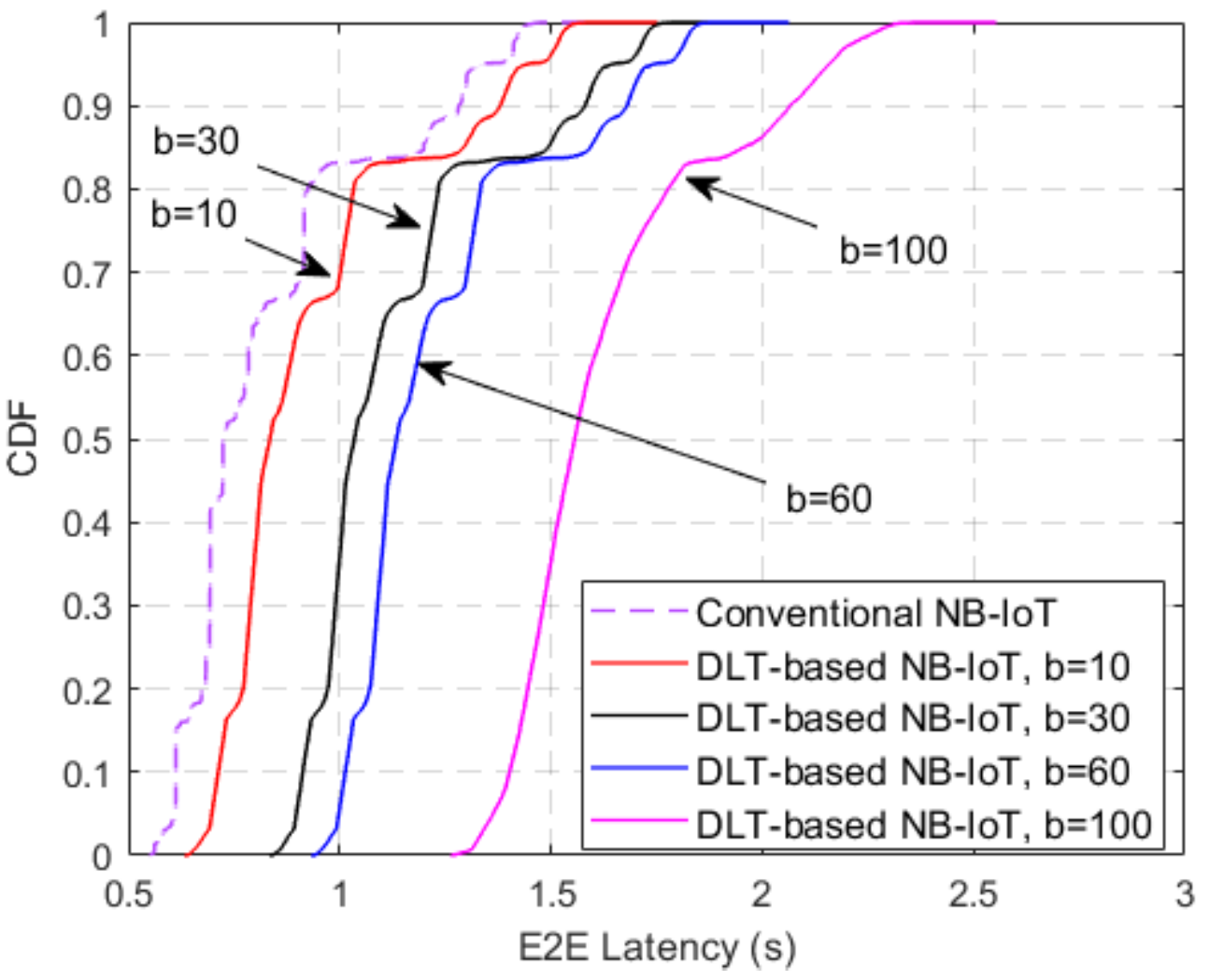}
    \caption{E2E latency of our Blockchain-enabled NB-IoT monitoring system. 
    }
    \label{fig:e2elatency}
\end{figure}

In these experiments, we configured two UEs, with one $\text{CO}_2$ sensor each, to gather data and upload to the ledger once every $10$~s. Our results are shown in Fig. \ref{fig:e2elatency}, where various block sizes $b$, from $10$ to $100$~transactions per block, were considered; the E2E latency for traditional NB-IoT packets is included as a reference.



As expected, Fig. \ref{fig:e2elatency} shows that an increase in block size comes with a slight increase in E2E latency. However, this increase is minor, even when compared to conventional NB-IoT packets, and may be suitable for most IoT monitoring applications. Specifically, the E2E latency is doubled from an average $0.832$~s for conventional NB-IoT to $1.63$~s for DLT-based NB-IoT with block size $b=100$ transactions per block. Naturally, smaller block sizes lead to a smaller E2E latency, which is comparable to that of conventional NB-IoT, especially for $b=10$. The reason is that the block creation time in Hyperledger Fabric increases with the increase of the block size.  On the other hand, it is advisable to use small block sizes in monitoring applications where even conventional NB-IoT is close to the upper limit of the acceptable E2E latency of the system.


\section{Conclusion}
\label{sec:conclusion}
\PP{Monitoring of emissions based with IoT devices requires sets high demands for data reliability and trustworthiness.  A promising approach in that direction is integration of a Blockchain into the IoT system.
We have implemented Blockchain into an environmental monitoring system based on NB-IoT, analyzed the tradeoffs and evaluated the performance.Our results show that the integration of Blockchain increases the load in the downlink (DL) channel of NB-IoT, unlike the plain variant of NB-IoT that does not use Blockchain.} 
Furthermore, both the level of security and the DL traffic load increase with the number of endorsing peers in Hyperledger. Besides, the E2E latency of the monitoring system increases slightly with the block size. This behavior was expected, but our results show that the increase in E2E latency is small even when compared to conventional NB-IoT. Therefore, integrated Blockchain and NB-IoT monitoring systems provide valuable benefits to a wide range of environmental applications and, in particular, to those that deal with sensitive information, such as carbon emissions or air pollution monitoring.

\section*{Acknowledgment}
This work has been in part supported by the European Research Council (Horizon 2020 ERC Consolidator Grant Nr. 648382 WILLOW).

\bibliographystyle{IEEEtran}
\bibliography{IEEEabrv,bibliography}

\section*{Biography}
\footnotesize
    \textbf{Lam D. Nguyen} (S'20) received the B.Sc. degree (Hons.) in electronics and telecommunications from the Hanoi University of Science and Technology, Hanoi, Vietnam, in 2015, and the M.Sc. degree in Computer Science from the Seoul National University, South Korea, in 2018. He is currently pursuing the Ph.D. degree Aalborg University. His research interests lie at the intersection of operations research, Blockchain, Internet of Things.

    \textbf{Anders E. Kal{\o}r} (S'17) received the B.Sc. degree in computer engineering and the M.Sc. degree in networks and distributed systems from Aalborg University, Denmark, in 2015 and 2017, respectively. He is currently pursuing a Ph.D. degree in the area of wireless communications and networking at Aalborg University. His research interests include communication theory, MAC layer design for wireless systems, and networking.

    \textbf{Israel Leyva-Mayorga} (M'20) received the M.Sc. degree (Hons.) in mobile computing systems from the Ins\-ti\-tu\-to Po\-li\-t\'ec\-ni\-co Na\-cio\-nal (IPN), Mexico, in 2014 and the Ph.D. (\emph{Cum Laude}) in telecommunications from the U\-ni\-ver\-si\-tat Po\-li\-t\`ec\-ni\-ca de Va\-l\`en\-cia (UPV), Spain, in 2018. He is currently a postdoc at the Department of Electronic Systems, Aalborg University (AAU), Denmark.

    \textbf{Petar Popovski} (S'97, A'98, M'04, SM'10, F'16) is a professor at Aalborg University, where he heads the Connectivity Section. He received his Dipl.-Ing./ Magister Ing. in communication engineering from Sts. Cyril and Methodius University in Skopje and his Ph.D. from Aalborg University. He received an ERC Consolidator Grant (2015) and the Danish Elite Researcher award (2016). He is an Area Editor for IEEE Transactions on Wireless Communications. His research interests are in wireless communications/networks and communication theory.


\end{document}